\documentclass[11pt]{article}

\usepackage{acl}
\usepackage{times}
\usepackage{latexsym}
\usepackage[T1]{fontenc}
\usepackage[utf8]{inputenc}
\usepackage{microtype}
\usepackage{inconsolata}
\usepackage{graphicx}
\usepackage{booktabs}
\usepackage{amsmath}
\usepackage{amssymb}
\usepackage{enumitem}

\title{Borrowed Strength, Imported Risk:\\
Capability Gaps and Reception Govern Heterogeneous LLM Debate}

\author{
\textbf{Prashanti Nilayam} \quad
\textbf{Kiran Kumar Ramanna} \quad
\textbf{Prashil Tumbade} \\
\textbf{Nagaraj N Bhat} \quad \textbf{Sankalp Nayak} \\
ServiceNow \\
{\small\texttt{\{prashanti.nilayam,kiran.ramanna,prashil.tumbade,}} \\
{\small\texttt{nagaraj.bhat,sankalp.nayak\}@servicenow.com}}
}

\begin{document}
\maketitle

\begin{abstract}
Heterogeneous LLM debate is often attributed to collective reasoning that creates
capability beyond any member. In the matched panels we study, the aggregate gain is dominated by
\emph{capability transfer}: a peer supplies task competence and receivers
absorb a measurable fraction. Across nine matched directions, honest benefit and
first-exchange adversarial exposure open when the peer has a positive same-task
advantage, and honest gain is closely described by the empirical decomposition
$\Delta A_d^{\mathrm{hon}}\approx\kappa\Gamma_d^+\rho_d$, the product of that
advantage and the receiver's reception. Two fixed-receiver directions provide
held-out validation without refitting: reception remains stable within uncertainty as a larger
peer advantage produces a larger gain. A matched, single-direction open-answer
MuSiQue scope probe reaches the negative-gap side of the same gate. Qwen-7B trails the
Gemma-3-12B receivers by 10.78 points; despite 62.50\% reception, it produces only
+1.17 points of controlled first-exchange gain. Reception cannot substitute for
positive aggregate peer advantage. When an influential peer is assigned an adversarial role, the
same influential peer slot removes the borrowed gain and can push the panel below homogeneous
debate. Across all five full-reference settings, the best observed majority correctness
does not exceed the influential member's best observed accuracy, despite unused complementary answers. Heterogeneous
debate is therefore a measurable transfer decision with a measurable failure
mode, not a benefit conferred by diversity alone.
\end{abstract}

\section{Introduction}
\label{sec:introduction}

Heterogeneous debate puts different language models in one panel because their
strengths are expected to compose. If accuracy rises, the standard explanation
is diversity: members contribute complementary reasoning and the exchange creates
an answer that no member could produce alone
\citep{du2024improving,yin2023exchange,wang2025mixture}. That explanation carries
an operational promise. A system that creates new collective capability may
justify several times the inference cost of calling one model; a system that
relays one capable member's answer must be evaluated against calling or routing
to that member directly. Recent evaluations increasingly favour the second
reading: debate often loses to single-agent inference at a fraction of the cost
\citep{zhang2025stop,bertalanic2026cost}. In the Bayesian model of
\citet{choi2025debate}, expected correctness is unchanged unless updates are
biased toward correct signals. What that work does not supply is the source of
that bias or a way to measure it in advance, which is what we provide.

We separate these accounts. Under \emph{capability composition}, the panel turns
distributed partial knowledge into capability beyond its members. Under
\emph{capability transfer}, one peer already has the task competence and the
debate routes some of it to weaker receivers. Our claim is that transfer dominates
net gain in these evaluated panels, not that item-level composition is absent.
Transfer makes four concrete
predictions: benefit should require something for the peer to supply; the amount
transferred should depend on the receiver's willingness to use a correct signal;
realized consensus should approach the influential peer rather than an oracle
over all answers; and changing that peer's role should reverse the channel that
delivered the gain.

We test all four predictions with matched three-agent panels. For each
receiver--peer--task direction, we compare receiver-homogeneous,
peer-homogeneous, honest heterogeneous, and adversarial heterogeneous conditions.
The peer's capability is measured on the same items under the same protocol.
Ordinary self-revision is absorbed by the receiver-homogeneous panel. Honest and
adversarial mixed panels hold model identity and position fixed and change one
role instruction. Nine directions span open and frontier models on
MATH-hard and GSM8K. Two fixed-receiver directions hold Qwen-7B fixed while replacing a
Gemma-3-12B peer with GPT-OSS-120B, a direct confirmatory test of whether a larger
aggregate peer advantage produces more gain at stable reception.

The data follow a two-factor decomposition. Let $\Gamma_d$ be the peer's
same-task accuracy advantage and $\rho_d$ the probability that a wrong receiver
becomes correct when the honest peer begins correct. Then
\begin{equation}
\Delta A_d^{\mathrm{hon}} \approx \kappa\Gamma_d^+\rho_d.
\label{eq:intro-product}
\end{equation}
The product fit has $R^2=.995$, against $.776$ for capability gap
alone. More importantly, the frozen model predicts both fixed-receiver
gains: reception remains near 61\% while doubling the peer gap raises controlled
R1 gain from 7.89 to 16.92 points.

A matched MuSiQue-Ans cohort provides a single-direction scope probe of the
negative-gap regime outside mathematical reasoning. It compares Gemma-3-12B
receivers with a Qwen-7B peer on 300 open-answer multi-hop
questions, including homogeneous, honest, and adversarial panels. Qwen is 10.78
points less accurate than Gemma at R0. The receivers nevertheless receive its
correct answers at 62.50\%, yet controlled first-exchange gain is only +1.17
points. The scope probe makes the distinction clear: reception can be high, but
a negative aggregate peer advantage leaves little net gain.

The same design exposes the cost of transfer. Across the nine directions,
reversing the peer's role removes recovery, raises corruption, and leaves the
adversarial panels 3.06 points below homogeneous debate at R1. In the fixed-receiver
GPT-OSS panel, the R4 swing is $-27.60$ points. Borrowed accuracy is
surrenderable because the component that supplies it also holds leverage over
the panel.

Our contributions are:
\begin{enumerate}[leftmargin=1.2em,itemsep=1pt,topsep=2pt]
  \item A matched measurement that tests whether aggregate gains reflect
  capability transfer or composition using homogeneous, heterogeneous, and member-level
  reference panels.
  \item A \emph{capability gate} across the mathematical-reasoning directions:
  net benefit and first-exchange leverage remain small without positive measured
  peer advantage, with a single-direction open-answer multi-hop QA scope probe.
  \item \emph{Reception power} and an advantage--reception decomposition whose frozen coefficients predict fixed-receiver gains without refitting.
  \item A benefit--risk account of borrowed strength: role reversal can consume
  the gain and push the panel below homogeneous debate, while consensus leaves
  available oracle complementarity unrealized.
\end{enumerate}

\section{Framework}
\label{sec:framework}

\paragraph{Directions and panels.}
A direction $d=(R,P,\mathcal{T})$ fixes a receiver model, peer model, and task.
We compare four matched panels:
\begin{center}
\small
\begin{tabular}{ll}
$[R,R,R]$ & receiver-homogeneous control \\
$[P,P,P]$ & peer-homogeneous capability reference \\
$[R,R,P]$ & honest heterogeneous panel \\
$[R,R,P^{\mathrm{adv}}]$ & same peer with an adversarial role \\
\end{tabular}
\end{center}
Slots 0--1 are receivers and slot 2 is the peer. Only the role instruction
changes between the two mixed panels.

\paragraph{Transitions.}
Let $C_{dri}^{x}\in\{0,1\}$ denote semantic correctness for receiver exposure
$i$ at round $r$ in condition $x\in\{\mathrm{hom},\mathrm{hon},\mathrm{adv}\}$.
The primary estimand is the first exchange, R0$\rightarrow$R1:
\begin{align}
W_d^x &= P(C_{d1}^{x}=1\mid C_{d0}^{x}=0),\\
Q_d^x &= P(C_{d1}^{x}=0\mid C_{d0}^{x}=1),\\
\Delta A_d^x &= [A_{d1}^x-A_{d0}^x]-[A_{d1}^{\mathrm{hom}}-A_{d0}^{\mathrm{hom}}].
\end{align}
$W_d^x$ and $Q_d^x$ are recovery and corruption. ``Honest'' labels the
non-adversarial panel; controlled honest gain subtracts homogeneous revision
from the mixed panel's accuracy change. Every reported aggregate is
an equal-weight mean over named directions; item counts never determine a
direction's weight.

\paragraph{Aggregate advantage and headroom.}
On each analysis's matched support,
\begin{align}
\Gamma_d &= A_{d0}^{[P,P,P]}-A_{d0}^{[R,R,R]}, &
H_d &= 1-A_{d0}^{[R,R,R]},
\end{align}
and $\Gamma_d^+=\max(0,\Gamma_d)$. $\Gamma_d$ measures net aggregate peer
advantage, not the total mass of items on which only the peer is correct;
$H_d$ bounds the receiver's possible improvement.

\paragraph{Reception and adversarial uptake.}
Let $C_{d0}^{P}$ denote the heterogeneous peer's R0 correctness. We define
\begin{align}
\rho_d &=P(C_{d1}^{\mathrm{hon}}=1\mid C_{d0}^{\mathrm{hon}}=0,C_{d0}^{P}=1),\\
\rho_{0d}&=P(C_{d1}^{\mathrm{hon}}=1\mid C_{d0}^{\mathrm{hon}}=0,C_{d0}^{P}=0),\\
\upsilon_d&=P(C_{d1}^{\mathrm{adv}}=0\mid C_{d0}^{\mathrm{adv}}=1,C_{d0}^{P}=0).
\end{align}
$\rho_d$ is reception power; $\rho_{0d}$ measures revision without a correct
peer signal; $\upsilon_d$ is adversarial uptake.

For example, in the Qwen-7B$\leftarrow$Gemma-3-12B direction, receiver R0 accuracy is 58.7\%, peer
accuracy 67.8\%, $\Gamma=+9.1$ points, and headroom 41.3 points. Among initially
receiver-wrong/peer-correct cases, reception is $\rho=62.2\%$ (95\% CI
[50.0,73.7]); controlled honest gain is +7.89 points. The observed gain need not
equal $\Gamma\rho$; Eq.~\ref{eq:product} includes fitted scaling and a residual term.

\paragraph{Predictions.}
Capability transfer yields four testable expectations:
\begin{enumerate}[leftmargin=1.2em,itemsep=1pt,topsep=2pt]
\item The marginal honest gain and first-exchange adversarial leverage of an
added heterogeneous peer remain small when $\Gamma_d\leq0$.
\item Honest gain is approximately
\begin{equation}
\Delta A_d^{\mathrm{hon}}=\kappa\Gamma_d^+\rho_d+\varepsilon_d,
\label{eq:product}
\end{equation}
where $\varepsilon_d$ collects unaided revision, non-nested correctness, and
finite-sample error. Reception and gain share R1 outcomes; under near-nested
correctness, eligible correction mass approaches $\Gamma_d$. Thus the product
fit is partly structurally induced, not independent evidence; the cross-peer
predictions provide a test beyond the in-sample fit.
\item Once a peer supplies material gain, reversing its role removes some or all
of that borrowed gain and may create net loss relative to $[R,R,R]$.
\item Realized panel accuracy approaches the influential peer rather than the
oracle over all member answers.
\end{enumerate}

Section~\ref{sec:gate} tests $\Gamma$, Section~\ref{sec:reception} examines
$\rho$ and the product fit, and Section~\ref{sec:reversal} examines $\upsilon$
and role reversal. Practical use is discussed in Section~\ref{sec:discussion},
``The relevant baseline is the influential member.''

\section{Experimental Setup}
\label{sec:experiments}

\paragraph{Primary directions.}
The primary analysis contains nine directions (Table~\ref{tab:directions}): six
on the 262 level-4/5 items of MATH-500 (MATH-hard)
\citep{hendrycks2021measuring,lightman2024let} and three on GSM8K
\citep{cobbe2021training}. It spans 1B--120B open models and three frontier APIs.
GPT-4.1/Gemini-Flash and Llama-8B/Qwen-7B are evaluated in both orientations,
so the sign of $\Gamma_d$ is not fixed by model identity.\footnote{Reproduction code and artifact documentation:
\url{https://github.com/prashanti-nilayam/BSIR}.}

\paragraph{Confirmatory directions.}
Two further MATH-hard directions, specified after Eq.~\ref{eq:product}, hold Qwen-7B fixed as $R$ and either Gemma-3-12B or GPT-OSS-120B as
$P$. The same Qwen homogeneous panel is the control. This design holds dataset,
receiver, headroom, slots, and protocol fixed while moving the measured peer gap
from 9.1 to 21.5 points. They were not used to formulate the approximation, and we report them separately from the pooled sensitivity.
An additional same-family Qwen-72B honest peer provides a control for the fixed
Qwen-7B receiver; this condition is excluded from the original fits and held-out
predictions.

\paragraph{Cross-task scope probe.}
On 300 MuSiQue-Ans open-answer multi-hop questions
\citep{trivedi2022musique}, Gemma-3-12B is the receiver and Qwen-7B the peer. We
run $[G,G,G]$, $[Q,Q,Q]$, $[G,G,Q]$, and $[G,G,Q^{\mathrm{adv}}]$ on the same
items, plus a same-family adversary $[G,G,G^{\mathrm{adv}}]$ and a contaminated
panel $[G,G^{\mathrm{adv}},Q]$. This cohort tests the negative-gap prediction on
a non-mathematical task; it is not pooled into the cross-direction correlations or the fixed-receiver test.

\begin{table*}[t]
\centering
\small
\begin{tabular}{llrrrrr}
\toprule
Direction ($R\leftarrow P$) & Task & Rep.-items & $A_{0,R}$ & $A_{0,P}$ & $\Gamma_d$ & $H_d$ \\
\midrule
Gemma-3-12B $\leftarrow$ Qwen-7B & MATH-hard & 786 & 67.8 & 58.1 & $-$9.7 & 32.2 \\
Qwen-7B $\leftarrow$ Llama-8B & GSM8K & 3957 & 91.9 & 87.1 & $-$4.8 & 8.1 \\
Gemini-Flash $\leftarrow$ GPT-4.1 & MATH-hard & 786 & 79.4 & 75.9 & $-$3.5 & 20.6 \\
GPT-4.1 $\leftarrow$ Gemini-Flash & MATH-hard & 786 & 75.9 & 79.4 & +3.5 & 24.1 \\
GPT-4.1 $\leftarrow$ OSS-120B & MATH-hard & 786 & 75.9 & 80.2 & +4.3 & 24.1 \\
Llama-8B $\leftarrow$ Qwen-7B & GSM8K & 3957 & 87.1 & 91.9 & +4.8 & 12.9 \\
Qwen-72B $\leftarrow$ OSS-120B & MATH-hard & 773 & 66.5 & 81.2 & +14.7 & 33.5 \\
Llama-1B $\leftarrow$ Gemma-2B & GSM8K$^\dagger$ & 100 & 37.7 & 61.0 & +23.3 & 62.3 \\
Llama-70B $\leftarrow$ OSS-120B & MATH-hard & 786 & 44.2 & 80.2 & +36.0 & 55.8 \\
\midrule
Qwen-7B $\leftarrow$ Gemma-3-12B & MATH-hard$^\ddagger$ & 786 & 58.7 & 67.8 & +9.1 & 41.3 \\
Qwen-7B $\leftarrow$ OSS-120B & MATH-hard$^\ddagger$ & 786 & 58.7 & 80.2 & +21.5 & 41.3 \\
\bottomrule
\end{tabular}
\caption{Evaluated directions on matched R0/R1 support. The first nine form the
nine-direction support (12{,}717 replicate-items); $\ddagger$ marks the two
fixed-receiver confirmatory directions (1{,}572 additional replicate-items).
$A_{0,R}$ and $A_{0,P}$ are homogeneous-panel accuracies in percent,
$\Gamma_d=A_{0,P}-A_{0,R}$, and $H_d=100-A_{0,R}$. $\dagger$ is a one-replicate,
100-item diagnostic; all others use three replicates.}
\label{tab:directions}
\end{table*}

\paragraph{Protocol.}
Following \citet{du2024improving}, three agents answer independently at R0 and
revise synchronously for four rounds after seeing the other two responses in
full. Primary runs use temperature 0 and three seeds/replicates except the
100-item small-model diagnostic. MATH-hard uses a 4,096-token completion budget;
GSM8K uses 1,024. MuSiQue uses one temperature-0 replicate on 300 items. The
adversarial peer receives a committed plausible-wrong
target and is instructed to defend it. No other panel component changes.

\paragraph{Matching and scoring.}
We retain replicate-items containing all R0/R1 cells required by the four-panel
contrast. Last-record-wins indexing resolves resumed-run duplicates. Answers
are reparsed with benchmark-specific semantic equivalence; missing or
unparseable answers count as incorrect. The nine-direction support contains 12{,}717 replicate-items and 76{,}302 receiver-condition observations; the
two confirmatory directions add 1{,}572 replicate-items and 9{,}432 observations.
Every MuSiQue condition contains all 300 matched items, three agents, and five
rounds. MuSiQue answers are scored by alias-normalized exact match and token F1
without an LLM judge.
Later-round trajectories require complete R0--R4 support and are reported
separately from the first-exchange estimand.

\paragraph{Inference.}
Direction-level means are tested by exact sign flips. Cross-direction
associations use exact label permutations and leave-one-direction-out refits.
Intervals use a global benchmark-item Poisson bootstrap: an item's weight is
shared across every replicate, slot, panel, and direction in which it appears.
Logistic sensitivities use direction fixed effects and direction-clustered
sandwich errors. The confirmatory fixed-receiver contrasts use paired
benchmark-item bootstraps; their item-level trend model clusters by benchmark
item and controls replicate and receiver slot.

\paragraph{Diversity control.}
Appendix~\ref{app:controls} reports a same-model MATH-hard panel that varies prompts,
temperatures $(0,.3,.7)$, and seeds $(0,1,2)$ to elicit complementary reasoning
without introducing a model-capability gap.

\section{Results}
\label{sec:results}

Corresponding bootstrap intervals and, where applicable, exact tests are
reported in the appendix tables---Tables~\ref{tab:role} and
\ref{tab:musique-heterogeneous} for the aggregate and cross-task contrasts. The
text quotes an interval only where its width carries the argument.

\subsection{Honest influence is corrective and capability-gated}
\label{sec:gate}

Across the nine directions, the equal-weight R0$\rightarrow$R1 mean
of $W_d^{\mathrm{hon}}-W_d^{\mathrm{hom}}$ is +20.40 percentage points
(exact $p=.0039$). The corresponding mean
$\Delta A_d^{\mathrm{hon}}$ is +6.55 points ($p=.0078$), while honest corruption
changes by $-0.78$ points ([$-2.28$,0.74], $p=.3555$). Thus the aggregate is driven by correction of initially wrong receivers, not systematic
destabilization of correct ones.

\begin{table}[t]
\centering
\small
\begin{tabular}{lrr}
\toprule
Quantity (pp) & Ungated & Gated \\
\midrule
Honest recovery & +7.33 (.250) & +26.94 (.031) \\
Honest net accuracy & +0.49 (.500) & +9.58 (.031) \\
Adv.\ corruption & +0.36 (.250) & +10.16 (.031) \\
Adv.\ net accuracy & +0.46 (.750) & $-$4.82 (.094) \\
\bottomrule
\end{tabular}
\caption{The capability gate. Means over directions, in percentage points against
matched homogeneous debate, with exact sign-flip $p$ in parentheses. The worst
single ungated direction shows $+0.5$~pp of excess corruption against
$+24.6$~pp for the worst gated one.}
\label{tab:gate}
\end{table}

\begin{figure*}[t]
\centering
\includegraphics[width=\textwidth]{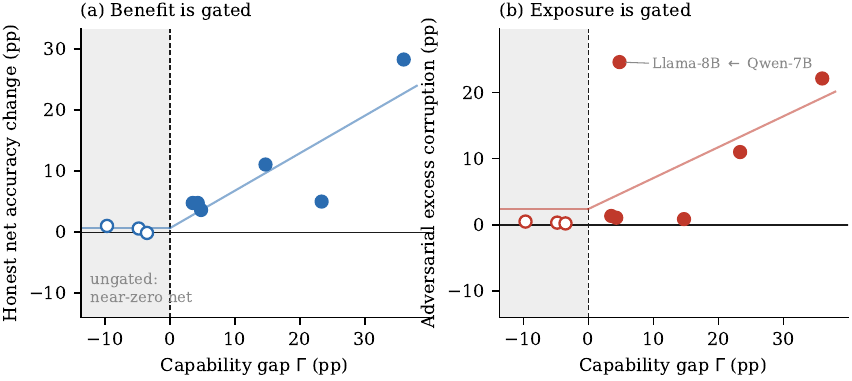}
\caption{The capability gate across nine directions. Each point is one
receiver--peer--task direction at R0$\rightarrow$R1. Left: controlled honest
gain. Right: adversarial excess corruption over the matched receiver-homogeneous
panel. Hollow points have $\Gamma_d<0$; filled points have $\Gamma_d>0$.}
\label{fig:gate}
\end{figure*}

The aggregate separates at $\Gamma_d=0$ (Figure~\ref{fig:gate}; Table~\ref{tab:gate}). The three ungated directions average +0.49 points of controlled honest accuracy;
the six gated directions average +9.58 points, positive in all six. Under an
adversarial role, excess corruption over $[R,R,R]$ averages +0.36 points in the
ungated group and +10.16 in the gated group. These are direction averages, not
pair-invariant effects. Across the nine directions, $\Gamma_d$ correlates with
honest gain at $r=.862$ (exact $p=.0033$) and with adversarial uptake at $r=.780$
($p=.0134$). The exact permutation tests operate on direction labels; because
several directions share models, tasks, and traces, they quantify association
within this evaluated set rather than among independent population draws.

\subsection{A cross-task scope probe reaches the negative-gap regime}
\label{sec:musique-gate}

The MuSiQue cohort changes the task and scoring regime while preserving the
matched panel design. Qwen-7B is 10.78 points less accurate than Gemma-3-12B at
R0. As an honest peer it produces only +1.17 points of controlled R1 gain. The small net effect is
not evidence that the receivers ignore Qwen. Reception is 62.50\%, against 4.91\% when Qwen is also wrong. Instead,
the honest peer raises recovery by 9.67 points and corruption by 9.25 points; the movements cancel in net accuracy.

The adversarial Qwen peer adds 2.66 points of first-exchange corruption over the
Gemma-homogeneous panel, with 4.85\% adversarial uptake. Thus a
non-mathematical, open-answer direction reaches the ungated regime: reception
can be active while net benefit and first-exchange heterogeneous leverage remain
small. Positive aggregate peer advantage, not diversity or reception alone, is the gate
(Table~\ref{tab:musique-heterogeneous}).

The two ungated regimes reach that null differently, and the difference is
informative. The three ungated mathematical directions combine a modest recovery
rise of 7.33 points with almost no corruption, 0.76 points; MuSiQue cancels two
large flows instead. What persists in this scope probe is the absence of net
benefit without positive aggregate peer advantage, not a single transition-level signature of it.

\subsection{Gain scales approximately with aggregate advantage times reception}
\label{sec:reception}

The product form follows the transition structure: when peer and receiver
correctness are nearly nested, positive capability gap tracks the cases in which
the receiver is wrong and the peer is correct. Outside that regime, $\Gamma_d$
is a net aggregate advantage; $\Gamma_d\leq0$ does not imply the absence of
useful peer-only answers (Appendix~\ref{app:sensitivity}). Reception measures the conversion
of available peer advantage; in the full nine-direction set, incorporating it
sharply reduces the gap-only residual, while both specifications fit closely in
the sensitivity analysis (Appendix~\ref{app:sensitivity}). Across the nine
directions, $\rho_d$ ranges from 14.7\% to
83.3\% and exceeds its unaided control
$\rho_{0d}$ in every direction (Appendix Figure~\ref{fig:reception}, left). Neither $\Gamma_d$ nor $H_d$ significantly
predicts reception ($r=.251$, $p=.5157$ and $r=-.238$, $p=.5230$). A hinge on
$\Gamma_d^+$ fits controlled gain with $R^2=.776$ and a 9.97-point maximum
residual. Replacing it with $\Gamma_d^+\rho_d$ yields slope 1.088,
$R^2=.995$, and a 1.07-point maximum residual (Appendix Figure~\ref{fig:reception}, right; Table~\ref{tab:specifications}).

\begin{table}[t]
\centering
\small
\begin{tabular}{lrrrl}
\toprule
Predictor & Slope & $R^2$ & Exact $p$ & LOO slope \\
\midrule
$\Gamma^{+}$ & 0.614 & 0.7760 & 0.0078 & [0.28, 0.76] \\
$\rho$ & 11.971 & 0.1455 & 0.3560 & [5.46, 15.39] \\
$\Gamma^{+} \times \rho$ & 1.088 & 0.9950 & 0.0002 & [1.08, 1.12] \\
\bottomrule
\end{tabular}
\caption{Cross-direction models of the honest net accuracy gain (pp), fitted on
identical inputs over the same nine directions. $\Gamma^{+} = \max(0, \Gamma)$
encodes the gate. Exact $p$ enumerates all $9! = 362{,}880$ label permutations;
\emph{LOO slope} is the range over the nine leave-one-direction-out refits.}
\label{tab:specifications}
\end{table}

\subsection{A fixed receiver provides held-out validation}
\label{sec:confirmation}

\begin{figure*}[t]
\centering
\includegraphics[width=\textwidth]{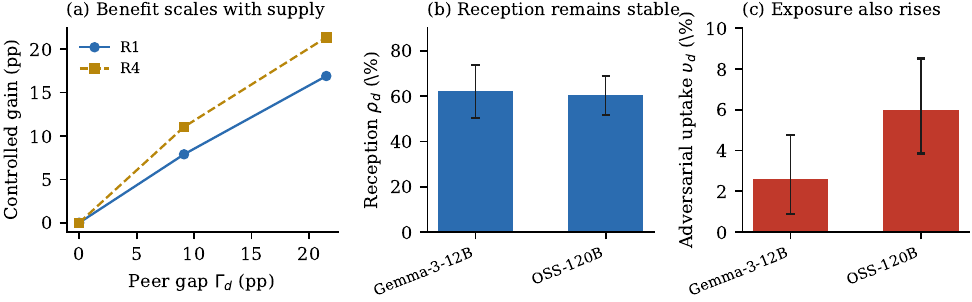}
\caption{Confirmatory fixed-receiver experiment on MATH-hard. Left: controlled gain
increases with peer capability gap at both R1 and R4. Middle: reception remains
stable across the two heterogeneous peers. Right: adversarial uptake is higher
with GPT-OSS than with Gemma under the same adversarial role. Intervals are
paired benchmark-item bootstraps.}
\label{fig:confirmatory}
\end{figure*}

The two fixed-receiver directions provide a controlled held-out test. Replacing Gemma-3-12B
with GPT-OSS-120B raises $\Gamma_d$ from 9.12 to 21.50 points, while Qwen's
measured reception remains effectively unchanged (Figure~\ref{fig:confirmatory}, middle): 62.16\% versus 60.35\%, a
$-1.81$-point difference with paired bootstrap CI [$-13.01$,9.87]
($p=.7875$). Controlled R1 gain nevertheless rises from +7.89 to +16.92 points (Figure~\ref{fig:confirmatory}, left);
the paired difference is +9.03 [5.14,12.88] ($p<.0005$).

\begin{table}[t]
\centering
\small
\setlength{\tabcolsep}{4pt}
\begin{tabular}{lrrr}
\toprule
Qwen-7B peer & Qwen & Gemma & OSS-120B \\
\midrule
$\Gamma_d$ (pp) & 0.00 & 9.12 & 21.50 \\
$\rho_d$ (\%) & -- & 62.16 & 60.35 \\
R1 gain (pp) & 0.00 & +7.89 & +16.92 \\
R4 gain (pp) & 0.00 & +11.05 & +21.37 \\
R1 recovery (\%) & 2.92 & 25.26 & 42.94 \\
R1 honest corruption (\%) & 2.06 & 4.72 & 2.21 \\
$\upsilon_d$ (\%) & -- & 2.60 & 6.00 \\
\bottomrule
\end{tabular}
\caption{Fixed-receiver MATH-hard ladder. Gain is the receiver's controlled accuracy
change relative to $[Q,Q,Q]$ at the indicated round. Reception and adversarial
uptake follow \S\ref{sec:framework}. The zero-gap column is the homogeneous
reference, not a separately designated honest peer.}
\label{tab:confirmatory}
\end{table}

Using coefficients fitted on the nine directions, without refitting, reception measured with
Gemma predicts +15.14 points for Qwen with GPT-OSS; the observed gain is +16.92.
Conversely, reception measured with GPT-OSS predicts +6.59 points for the Gemma
direction; the observed gain is +7.89. Across the zero-gap Qwen control and the
two heterogeneous peers, gain is nearly linear in gap at R1 (through-origin
slope .799, $R^2=.999$) and R4 (slope 1.027, $R^2=.994$; Table~\ref{tab:confirmatory}). At the item level,
a ten-point increase in peer gap multiplies recovery odds by 3.03
([2.47,3.73]), while honest corruption has no gap trend (OR .98 [.72,1.34]).
These item-level intervals quantify these 262 items; the three-point ladder is
the model-level evidence.

With the GPT-OSS direction included, the empirical pattern persists:
the gap--gain correlation across the resulting ten directions becomes $r=.876$ (exact $p=.00080$), the
gap--reception association remains small and nonsignificant ($r=.263$,
$p=.459$), and the product fit remains
$R^2=.990$ against $.795$ for the hinge. With both included, $R^2=.989$ for the product and $.793$ for the hinge. These pooled refits
are sensitivity analyses.

The additional same-family Qwen-7B$\leftarrow$Qwen-72B control on MATH-hard
(three replicates, 786 replicate-items) has $\Gamma=+7.1$ points,
$\rho=52.2\%$ [38.5,65.8], and controlled R1 gain +3.4 points. Its reception
interval overlaps that of the approximately gap-matched Gemma peer, revealing
no clear family-aligned difference; this does not establish equality or remove
identity/style confounding. Its reception denominator is 180. The control has
no adversarial arm and is excluded from the original fits.

\subsection{The same influential peer slot imports risk}
\label{sec:reversal}

In the nine directions, changing the peer from honest to adversarial
reduces recovery by an equal-weight mean of 15.25 points (exact $p=.0156$) and increases corruption by 7.67 points ($p=.0312$). Consequently, the adversarial mixed panels average
3.06 points below their matched homogeneous controls at R1 (Appendix Figure~\ref{fig:surrender}, left; Appendix~\ref{app:rates}). The fixed-receiver
comparison points in the same direction: adversarial uptake rises from 2.60\%
with Gemma to 6.00\% with GPT-OSS (Figure~\ref{fig:confirmatory}, right), a paired difference of +3.41 points
([1.12,5.82], $p=.0040$). Compliance differs across peers: 68.54\% for Gemma
and 99.75\% for GPT-OSS. At R4, Gemma's honest-to-adversarial swing is
$-18.96$ points by intent to treat and $-24.72$ per protocol (adversary
parsed-wrong at R0--R4). For GPT-OSS,
Qwen with an honest peer is +21.37 points above homogeneous debate; assigning
that peer the adversarial role produces a $-27.60$-point intent-to-treat swing
($-27.93$ per protocol) and leaves Qwen 6.23 points below homogeneous debate.
The GPT-OSS swing is therefore robust, while its difference from Gemma cannot
be attributed to capability alone.

Correction and corruption are not one generic credulity trait. Across the nine directions, $\rho_d$ does not predict $\upsilon_d$ ($r=.223$, $p=.554$), whereas the
capability gap does. With GPT-OSS added, the latter association weakens but
remains significant ($r=.683$, exact $p=.0282$). We therefore interpret
$\Gamma_d$ as a recurring risk predictor, not a complete model of adversarial
loss.

MuSiQue also locates the boundary of this claim. Repeated exposure to the weaker
adversarial Qwen peer accumulates: by R4 its controlled effect is $-4.50$ points, an interval excluding zero, even though its R1 excess-corruption interval includes zero.
And a same-family Gemma adversary---a different threat model from adding a
heterogeneous peer---raises R1 corruption by 8.47 points and reaches a controlled R4 effect of $-9.50$ points. The capability gate
governs the marginal first-exchange leverage of a heterogeneous replacement; it
does not make repeated or same-family adversaries harmless.

\subsection{The panel approaches the influential member}
\label{sec:ceiling}

In the five full-reference settings, the panel's best majority correctness
(at least two of three agents correct) never exceeds the peer's best accuracy inside that
panel, while an oracle that credits any correct member is 4.7--23.0 points
higher (Appendix Figure~\ref{fig:surrender}, right; Table~\ref{tab:ceiling} in
Appendix~\ref{app:trajectories}). These finite-setting observations support a
borrowed ceiling; they are not a universal upper bound on heterogeneous
protocols, including routing or selection. At fixed R4, discarded correct answers exceed new correct outcomes
by 3.6--30 times in all five settings: transfer dominates net gain, while new
correct outcomes remain item-level exceptions to a pure relay account
(Appendix Table~\ref{tab:composition-accounting}).

Across the extended set's nine positive-gap directions (six primary, two
fixed-receiver, and the same-family control), five-round debate uses 40.6--86.7 times the total reported
tokens of one peer call (median 62.2). Even its retrospectively best majority
round beats the peer's single-pass R0 accuracy in only 3/9 directions. We charge
all 15 debate calls versus one peer call on matched items
(Appendix~\ref{app:runs}). This is token volume, not dollar cost or measured
latency; the protocol has five sequential stages. Strong-model resampling and
learned routing remain unevaluated.

\subsection{Robustness and controls}

Requiring the adversary to remain parsed and wrong in every round
strengthens the recovery reversal and preserves the corruption increase
(Appendix~\ref{app:rates}). Results retain their direction within MATH-hard and
GSM8K. The matched MuSiQue scope probe reaches the negative-gap side under
open-answer semantic scoring. The prompt/temperature/seed control creates more
complementary R0 answers but does not convert them into higher consensus accuracy
(Appendix~\ref{app:controls}).

\section{Discussion}
\label{sec:discussion}

\paragraph{In these panels, aggregate advantage gates net gain; reception sets the conversion rate.}
The experiments resolve two questions that panel accuracy collapses. In these tested panels,
material net gain is associated with positive aggregate advantage, measured by
$\Gamma_d$; this is not a necessary condition for item-level help. The receivers must then use the peer's
correct signal, measured by $\rho_d$.
Neither factor substitutes for the other. MuSiQue makes
the separation especially clear. The receivers receive a correct Qwen signal at
62.50\%, but Qwen is 10.78 points weaker than Gemma and controlled gain is only
+1.17 points. High reception without positive aggregate peer advantage creates
churn, not strength.

\paragraph{The fixed receiver provides held-out validation.}
The nine-direction fit alone would leave open whether
$\Gamma_d^+\rho_d$ merely summarizes nine directions. The fixed-receiver experiment
holds receiver, task, headroom, protocol, and homogeneous baseline constant. Reception estimated with one peer
predicts the gain with the other using coefficients frozen on the nine directions.
That cross-peer prediction provides held-out validation under a controlled
receiver, task, and protocol; it does not establish a universal relationship
across models, tasks, or debate protocols.

\paragraph{Borrowed strength has a replacement cost.}
The peer that supplies accuracy also controls whether that accuracy remains
available. Across the nine directions, role reversal consumes the honest gain and
leaves the adversarial panels below homogeneous debate. This is not a generic claim
that every weak peer is harmless or every capable peer is malicious. MuSiQue
shows the boundary directly: a weaker adversarial heterogeneous peer has little
first-exchange leverage, but repeated exposure eventually causes loss, while a
same-family committed adversary can cause immediate harm without a positive
heterogeneous capability gap. Operators therefore need both the marginal
heterogeneous-leverage measurement and an explicit threat model.

\paragraph{The relevant baseline is the influential member.}
Beating $[R,R,R]$ does not establish collective capability when $P$ can be called
directly. Heterogeneous transfer can still be valuable under rate limits, cost
constraints, or partial access to the peer. But the honest comparison is then
against strong-model inference and routing, not only against a weak homogeneous
panel. The oracle gap measures whether selection has complementarity left to
recover; the prompt/temperature/seed control shows that creating disagreement is
not enough when consensus revision discards it.

\paragraph{What the approximation means.}
Equation~\ref{eq:product} is a compact empirical account, not a new conservation
law. When peer and receiver correctness are nearly nested, the eligible mass on
which the receiver is wrong and the peer is right approaches $\Gamma_d$, making
part of the product structurally induced. Non-nested errors, unaided
revision, corruption, and sampling variation produce $\varepsilon_d$. The
evidentiary weight therefore rests on the residual reduction, the fixed-receiver
cross-peer predictions, and the negative-gap MuSiQue scope probe together---not on the nine-direction $R^2$ alone.

\section{Related Work}
\label{sec:related}

\paragraph{Debate and heterogeneous panels.}
Multi-agent debate can improve language-model reasoning
\citep{du2024improving,liang2024encouraging,yin2023exchange}, and heterogeneous
panels are often motivated by complementary perspectives
\citep{hegazy2024diversity,zhou2025adaptive,wang2025mixture}. The closest broad
evaluation is \citet{zhang2025stop}, who compare debate methods across models
and benchmarks, find frequent losses to simpler inference baselines, but identify
heterogeneity as a comparatively reliable remedy. We give that
remedy a measurable source--receiver structure: a fixed peer has a
same-task advantage $\Gamma_d$, a receiver has reception $\rho_d$, and later
fixed-receiver directions test their product.

Other work questions whether consensus debate creates new capability at all
\citep{smit2024should,yang2025revisiting,zhu2026demystifying,wu2025can,bertalanic2026cost}.
\citet{choi2025debate} show that majority voting accounts for most of the gain
usually credited to deliberation. In their Bayesian model, debate follows
martingale belief dynamics absent an update bias toward correct signals. A peer
with measured task advantage supplies such a signal, and $\rho_d$ measures the
fraction the receiver absorbs. The closest uncertainty decomposition is
\citet{qiao2026epistemic}, who separate inter-agent epistemic disagreement from
within-agent aleatoric uncertainty. Our decomposition instead separates peer-side
capability supply from receiver-side reception.
\citet{zhu2026demystifying} emphasize confidence and diversity;
\citet{wynn2025talk} show mixed-capability interaction can pull correct agents
toward incorrect answers. Selection- and routing-oriented systems instead
preserve disagreement and choose among members
\citep{cui2026freemad,verma2026selene,maryanskyy2026agents}. Our ceiling and
oracle measurements connect these strands: complementary answers are present in
our mathematical panels, but consensus revision does not realize them.

\citet{pappu2026teams} study expert underuse and consensus/robustness tradeoffs
in self-organizing teams. Our fixed composition, slots, and protocol instead
measure peer advantage, reception, and role reversal for a specified pair.
Their findings complement our ceiling and role-reversal results.

\paragraph{Persuasion, faulty peers, and revision.}
Adversarial or persuasive members can steer multi-agent systems
\citep{amayuelas2024multiagent,kraidia2026collaboration,agarwal2025persuasion,cui2025mad},
while Byzantine analyses characterize robustness primarily through the number
or placement of faulty agents
\citep{jo2025byzantine,zheng2026rethinking,tse2408resilience}.
\citet{nilayam2026borrowed} analyze receiver revision and contaminated-panel
robustness under honest and adversarial heterogeneous peers. We study a
different estimand: net capability transfer relative to homogeneous control,
and ask whether peer advantage and the receiver's reception predict its magnitude and
marginal adversarial leverage.

The closest transition-level comparison is \citet{qu2026easier}, who distinguish
beneficial revision from harmful revision using controlled peer messages and
find models easier to mislead than correct. Simulated messages give clean
control over consensus but have no supplying model whose task capability can be
measured. Our real-peer design adds that missing capability structure and finds
that beneficial reception exceeds adversarial uptake in most evaluated
directions. Work on sycophancy and stance convergence
\citep{sharma2024towards,ccelebi2025parrot,hao2026not} studies related response
changes, but generally begins from a correct model under social pressure.
Reception instead conditions on an initially wrong receiver and a correct peer.

\paragraph{Self-correction and compound-system propagation.}
Models often struggle to improve their own reasoning without external feedback
\citep{huang2023large}; consequently $[R,R,R]$ is an empirical control rather
than a presumed monotonic baseline. More broadly, work on error propagation in
compound systems separates the presence of an upstream signal from downstream
detection and correction \citep{nilayam2026detection}. Our
capability-supply/reception decomposition applies the same separation to
heterogeneous debate and adds an adversarial role reversal on the supplying
component.
QUIVER quantifies perturbation propagation and structural divergence in
graph-structured compound AI systems \citep{nilayam2026quiver}; here we measure
correctness transfer and adversarial influence under a fixed debate protocol.

\section{Conclusion}
\label{sec:conclusion}

In these panels, capability transfer dominates net gain, while item-level
composition is not ruled out. The peer's same-task advantage determines what can be supplied;
the receiver's reception determines how much is transferred. Nine directions expose the relationship, two fixed-receiver directions predict it without
refitting, and an open-answer MuSiQue scope probe reaches the negative-gap side:
in that direction, high reception produces movement but little net gain
from the inferior peer. The benefit is also
surrenderable. Reversing an influential peer's role consumes the borrowed gain
and can push the panel below homogeneous debate, while consensus still leaves
correct member answers unused. Heterogeneous debate should therefore be measured
as a supply, reception, and integrity problem---and compared with calling or
routing to the influential member directly.

\section*{Limitations}
\label{sec:limitations}

\paragraph{Finite and dependent directions.}
The cross-direction analysis contains nine primary and two fixed-receiver confirmatory directions. Several share models, tasks, and traces; they are not
independent draws from a population of model pairs. Exact direction-level tests
and item-clustered intervals quantify the evaluated comparisons without turning
them into population sampling.

\paragraph{Cross-task scope.}
The heterogeneous MuSiQue cohort provides a single-direction probe of the negative-gap gate on 300
open-answer multi-hop questions, but it is one model pair and one replicate. It
does not test positive-gap transfer or the product relationship outside
MATH-hard and GSM8K. Whether positive-gap transfer replicates on open-answer tasks is untested.

\paragraph{Empirical factors.}
The fixed-receiver ladder holds the receiver, task, protocol, and baseline constant,
but changing peer capability also changes peer identity and argument style.
Reception is measured under the debate protocol. We treat $\Gamma_d$ and
$\rho_d$ as separately measurable predictors, not intrinsic causal traits of a
model independent of its interaction context. Estimating them for a new
deployment requires a labeled, deployment-matched probe.

\paragraph{Structural coupling.}
When peer and receiver correctness are nested, the item mass underlying
$\rho_d$ approaches $\Gamma_d$. Equation~\ref{eq:product} therefore contains a structural
component in that regime. The fixed-receiver predictions and the MuSiQue
negative-gap result provide tests beyond the nine-direction fit, but do not remove the
structural component.

\paragraph{Threat model and rounds.}
The adversary follows one committed-wrong role prompt. The same-family MuSiQue
control demonstrates that lack of a positive heterogeneous gap does not protect
against every malicious participant, and the R4 trajectory shows that weak-peer
harm can accumulate after a small first exchange. Coordinated, adaptive, or
multi-peer attacks remain outside the evaluated threat model.

\paragraph{Protocol and ceiling.}
All panels use three agents, full transcript visibility, synchronous revision,
and one designated peer slot. Reception is an operational transition rate,
not a cognitive-pathway measure: answer-only, reasoning-only, and full-response
ablations are needed to distinguish adoption from reasoning-enabled correction. The borrowed-ceiling comparison selects the best
observed round in settings with complete standalone peer trajectories. It is a
property of these consensus panels, not an upper bound on routing, selection, or
other heterogeneous protocols.

\section*{Ethics Statement}
\label{sec:ethics}

This work studies a panel member instructed to defend a plausible incorrect
answer. The prompt is a direct role instruction rather than a jailbreak or a new
attack technique, and all adversarial runs concern public reasoning benchmarks.
We report it because identifying high-influence members as security-critical can
support safer deployment, auditing, and verification.

MATH, GSM8K, and MuSiQue are public datasets and contain no human-subject data
collected for this study. Open-weight models were served locally; hosted models
were accessed through their providers under standard terms. Analyses operate on
recorded traces and require no additional model calls. AI coding assistants were
used for analysis tooling and manuscript preparation; the authors verified the
reported quantities and remain responsible for the claims.

\bibliography{custom}

\appendix
\setcounter{figure}{\value{section}}
\renewcommand{\thefigure}{A\arabic{figure}}

\section{Reproducibility}
\label{app:repro}

The released artifact, documented at
\url{https://github.com/prashanti-nilayam/BSIR}, contains recorded JSONL traces, source manifests,
benchmark-aware parsers, analysis programs, and independent validators. The
validators reconstruct matched support and recount transition cells, bootstrap
intervals, direction-level tests, reception quantities, and ceiling statistics
without importing the analysis implementation. All manuscript figures and
tables are generated from frozen analysis outputs. Operational commands and
file-level hashes are provided in the artifact README rather than reproduced in
the paper.

\section{Run Configuration}
\label{app:runs}

Primary panels contain three agents and five synchronous rounds. MATH-hard uses
the 262 level-4/5 MATH-500 items and a 4,096-token completion budget; GSM8K uses
the full 1,319-item test set and 1,024 tokens. Primary decoding is temperature
0. The fixed-receiver MATH experiment uses three seeds and the same items, prompts,
rounds, and budgets in every condition. The Qwen homogeneous baseline is shared
by the Gemma and GPT-OSS directions. GPT-OSS follows the committed-wrong prompt
on 99.75\% of complete replicate-items across all five rounds; the corresponding
Gemma rate is 68.54\%, so intent-to-treat remains the primary comparison.
The MuSiQue scope cohort uses one temperature-0 replicate per condition on 300
matched items; every one of its six panels contains all 4,500 agent-round cells.

For the token comparison, matched replicate-items require complete five-round
honest panels and peer-reference R0 cells. We sum reported tokens across all
15 debate calls; the exchangeable peer-reference agents estimate the mean cost
and accuracy of one R0 call. This support can differ from the four-panel
first-exchange support. Different tokenizers make token volume an inference
proxy; best-round majority accuracy is selected retrospectively.

Models as served are GPT-4.1 \citep{openai2025gpt41}, Gemini-2.5-Flash
\citep{geminiteam2025gemini25}, GPT-OSS-120B \citep{openai2025gptoss},
Llama-3.1-70B-Instruct, Llama-3.1-8B-Instruct, and Llama-3.2-1B-Instruct
\citep{grattafiori2024llama3}; Qwen-2.5-72B-Instruct and Qwen-2.5-7B-Instruct
\citep{qwen2024qwen25}; and Gemma-3-12B-IT \citep{gemmateam2025gemma3} and
Gemma-2-2B-IT \citep{gemmateam2024gemma2}.

\section{Per-Direction Results}
\label{app:rates}

Table~\ref{tab:rates} reports per-direction recovery and corruption rates under
all three panel conditions, Table~\ref{tab:reception} the reception quantities
and their unaided controls, and Table~\ref{tab:wilson} Wilson intervals for the
underlying transition counts. Table~\ref{tab:role} aggregates the role effects at
first peer exposure and Table~\ref{tab:logistic} the logistic specifications.

\begin{table*}[t]
\centering
\small
\begin{tabular}{lrrrrrr}
\toprule
& \multicolumn{3}{c}{Recovery (\%)} & \multicolumn{3}{c}{Corruption (\%)} \\
\cmidrule(lr){2-4}\cmidrule(lr){5-7}
Direction & Homog. & Honest & Advers. & Homog. & Honest & Advers. \\
\midrule
Gemma-3-12B $\leftarrow$ Qwen-7B & 3.0 & 7.7 & 8.7 & 1.1 & 2.0 & 1.6 \\
Qwen-7B $\leftarrow$ Llama-8B & 3.9 & 17.7 & 13.1 & 0.1 & 0.6 & 0.4 \\
Gemini-Flash $\leftarrow$ GPT-4.1 & 17.1 & 20.6 & 15.2 & 1.5 & 2.4 & 1.8 \\
GPT-4.1 $\leftarrow$ Gemini-Flash & 26.6 & 41.8 & 30.7 & 8.8 & 6.5 & 10.2 \\
GPT-4.1 $\leftarrow$ OSS-120B & 26.6 & 41.7 & 26.8 & 8.8 & 7.3 & 9.9 \\
Llama-8B $\leftarrow$ Qwen-7B & 5.2 & 49.3 & 19.0 & 0.7 & 3.0 & 25.2 \\
Qwen-72B $\leftarrow$ OSS-120B & 2.7 & 35.2 & 5.4 & 1.5 & 1.0 & 2.3 \\
Llama-1B $\leftarrow$ Gemma-2B & 6.5 & 12.8 & 14.8 & 18.4 & 13.3 & 29.4 \\
Llama-70B $\leftarrow$ OSS-120B & 3.0 & 51.4 & 7.1 & 8.8 & 6.5 & 30.9 \\
\bottomrule
\end{tabular}
\caption{Per-direction R0$\to$R1 transition rates in all three matched conditions, intent to treat. The receiver-homogeneous column is the self-revision baseline that every honest and adversarial contrast in the main text is taken against.}
\label{tab:rates}
\end{table*}

\begin{table*}[t]
\centering
\small
\begin{tabular}{lrrlrrrr}
\toprule
Direction & $\Gamma$ (pp) & $\rho$ (\%) & Bootstrap 95\% CI & $n$ & Unaided (\%) & Uptake (\%) & Gain (pp) \\
\midrule
Gemma-3-12B $\leftarrow$ Qwen-7B & $-$9.7 & 14.7 & [2.6, 29.6] & 116 & 5.7 & 1.5 & +1.0 \\
Qwen-7B $\leftarrow$ Llama-8B & $-$4.8 & 36.9 & [25.2, 49.7] & 244 & 5.6 & 0.5 & +0.6 \\
Gemini-Flash $\leftarrow$ GPT-4.1 & $-$3.5 & 32.5 & [14.1, 51.6] & 114 & 13.9 & 1.8 & $-$0.1 \\
GPT-4.1 $\leftarrow$ Gemini-Flash & +3.5 & 83.3 & [72.2, 93.7] & 144 & 14.0 & 9.8 & +4.8 \\
GPT-4.1 $\leftarrow$ OSS-120B & +4.3 & 83.1 & [72.5, 92.7] & 154 & 13.3 & 9.9 & +4.8 \\
Llama-8B $\leftarrow$ Qwen-7B & +4.8 & 78.5 & [72.4, 84.2] & 632 & 2.3 & 25.2 & +3.6 \\
Qwen-72B $\leftarrow$ OSS-120B & +14.7 & 63.0 & [51.0, 74.7] & 254 & 7.8 & 2.3 & +11.1 \\
Llama-1B $\leftarrow$ Gemma-2B & +23.3 & 17.9 & [5.8, 31.2] & 56 & 8.2 & 29.4 & +5.0 \\
Llama-70B $\leftarrow$ OSS-120B & +36.0 & 71.0 & [62.9, 78.6] & 599 & 10.2 & 30.9 & +28.2 \\
\bottomrule
\end{tabular}
\caption{Reception power $\rho = P(\text{correct at R1} \mid \text{wrong at R0}, \text{honest peer correct at R0})$, with its control and its adversarial counterpart. \emph{Unaided} is the same conditional with the honest peer also wrong, and isolates ambient self-revision from peer-driven revision. \emph{Uptake} is $P(\text{wrong at R1} \mid \text{correct at R0}, \text{adversarial peer wrong at R0})$. \emph{Gain} is the honest net accuracy change relative to homogeneous debate. Intervals are global benchmark-item Poisson bootstraps; Wilson intervals are in the appendix.}
\label{tab:reception}
\end{table*}

\begin{table*}[t]
\centering
\small
\begin{tabular}{lrll}
\toprule
Direction & $\rho$ (\%) & Wilson 95\% & Bootstrap 95\% \\
\midrule
Gemma-3-12B $\leftarrow$ Qwen-7B & 14.7 & [9.4, 22.2] & [2.6, 29.6] \\
Qwen-7B $\leftarrow$ Llama-8B & 36.9 & [31.1, 43.1] & [25.2, 49.7] \\
Gemini-Flash $\leftarrow$ GPT-4.1 & 32.5 & [24.6, 41.5] & [14.1, 51.6] \\
GPT-4.1 $\leftarrow$ Gemini-Flash & 83.3 & [76.4, 88.5] & [72.2, 93.7] \\
GPT-4.1 $\leftarrow$ OSS-120B & 83.1 & [76.4, 88.2] & [72.5, 92.7] \\
Llama-8B $\leftarrow$ Qwen-7B & 78.5 & [75.1, 81.5] & [72.4, 84.2] \\
Qwen-72B $\leftarrow$ OSS-120B & 63.0 & [56.9, 68.7] & [51.0, 74.7] \\
Llama-1B $\leftarrow$ Gemma-2B & 17.9 & [10.0, 29.8] & [5.8, 31.2] \\
Llama-70B $\leftarrow$ OSS-120B & 71.0 & [67.2, 74.4] & [62.9, 78.6] \\
\bottomrule
\end{tabular}
\caption{Reception intervals under two schemes. Wilson intervals treat exposures as independent; the benchmark-item bootstrap resamples benchmark items and so absorbs clustering across replicates and receiver slots. The bootstrap intervals are wider everywhere and underlie the interval reporting in the manuscript.}
\label{tab:wilson}
\end{table*}

\begin{figure*}[t]
\centering
\includegraphics[width=\textwidth]{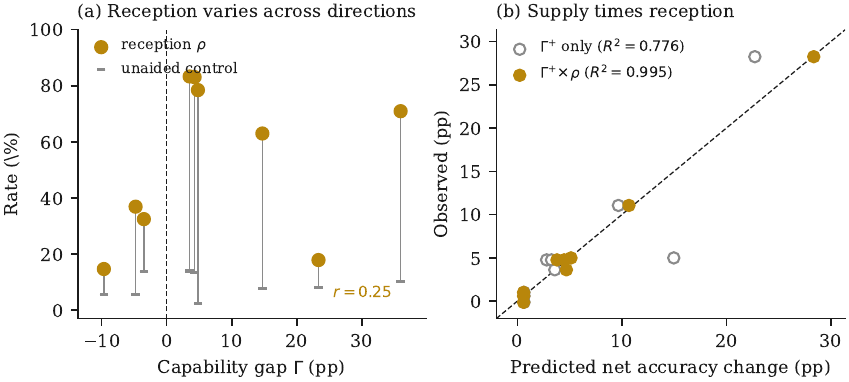}
\caption{Reception power in the nine directions. Left: $\rho_d$ and
its unaided control $\rho_{0d}$ against the measured gap. Right: observed gain
against predictions from the gap-only and product approximations. The fit is
descriptive over directions; the later fixed-receiver test is shown separately.}
\label{fig:reception}
\end{figure*}

\begin{table*}[t]
\centering
\small
\begin{tabular}{lrlrl}
\toprule
Contrast & Mean (pp) & Bootstrap 95\% CI & Exact $p$ & Odds ratio [95\% CI] \\
\midrule
\multicolumn{5}{l}{\emph{Intent to treat} (12{,}717 matched replicate-items)} \\
Recovery: honest vs.\ homogeneous & +20.40 & [+17.62, +23.28] & 0.0039 & 6.58 [2.24, 19.37] \\
Corruption: honest vs.\ homogeneous & $-$0.78 & [$-$2.28, +0.74] & 0.3555 & 1.43 [0.61, 3.36] \\
Recovery: adversarial vs.\ honest & $-$15.25 & [$-$17.94, $-$12.61] & 0.0156 & 0.29 [0.14, 0.62] \\
Corruption: adversarial vs.\ honest & +7.67 & [+5.98, +9.37] & 0.0312 & 5.17 [1.70, 15.72] \\
\midrule
\multicolumn{5}{l}{\emph{Per protocol} (10{,}580 items; adversary parsed-wrong in all of R0--R4)} \\
Recovery: honest vs.\ homogeneous & +20.08 & [+17.23, +23.08] & 0.0039 & -- \\
Corruption: honest vs.\ homogeneous & $-$0.22 & [$-$1.91, +1.46] & 0.7656 & -- \\
Recovery: adversarial vs.\ honest & $-$16.63 & [$-$19.53, $-$14.07] & 0.0078 & -- \\
Corruption: adversarial vs.\ honest & +7.77 & [+5.85, +9.70] & 0.0586 & -- \\
\bottomrule
\end{tabular}
\caption{Role effects at first peer exposure (R0$\to$R1), case-standardized over the nine directions. Exact $p$ is a paired sign-flip test over the $2^9$ direction-level sign assignments; intervals are global benchmark-item Poisson bootstraps (4{,}000 replicates). Odds ratios come from logistic models with a separate intercept per direction and direction-clustered sandwich standard errors on 8 degrees of freedom. Honest and adversarial contrasts use the same peer model in the same slot; only the role prompt changes.}
\label{tab:role}
\end{table*}

\begin{table*}[t]
\centering
\small
\begin{tabular}{lllrrlr}
\toprule
Outcome & Specification & Term & $n$ & OR & 95\% CI & $p$ \\
\midrule
recovery & pair fe three condition & honest vs.\ homogeneous & 14264 & 6.582 & [2.237, 19.367] & 0.0038 \\
recovery & pair fe three condition & adversarial vs.\ homogeneous & 14264 & 1.914 & [1.058, 3.464] & 0.0356 \\
corruption & pair fe three condition & honest vs.\ homogeneous & 62038 & 1.430 & [0.609, 3.357] & 0.3625 \\
corruption & pair fe three condition & adversarial vs.\ homogeneous & 62038 & 7.345 & [1.097, 49.158] & 0.0419 \\
recovery & pair fixed effects & adversarial vs.\ honest & 9493 & 0.293 & [0.137, 0.624] & 0.0057 \\
corruption & pair fixed effects & adversarial vs.\ honest & 41375 & 5.170 & [1.700, 15.725] & 0.0093 \\
recovery & gap & adversarial vs.\ honest & 9493 & 0.466 & [0.319, 0.680] & 0.0016 \\
recovery & gap & adversarial $\times$ $\Gamma$ & 9493 & 0.622 & [0.548, 0.706] & 0.0000 \\
corruption & gap & adversarial vs.\ honest & 41375 & 4.315 & [1.269, 14.673] & 0.0249 \\
corruption & gap & adversarial $\times$ $\Gamma$ & 41375 & 1.266 & [0.912, 1.757] & 0.1361 \\
\bottomrule
\end{tabular}
\caption{Logistic specifications, intent to treat. All models carry a separate intercept per direction and use direction-clustered sandwich standard errors with a $t$ reference on 8 degrees of freedom, so item count is never treated as the number of independent capability comparisons. The three-condition models include the receiver-homogeneous panel; $\Gamma$ is centred and scaled in percentage points.}
\label{tab:logistic}
\end{table*}

\begin{figure*}[t]
\centering
\includegraphics[width=\textwidth]{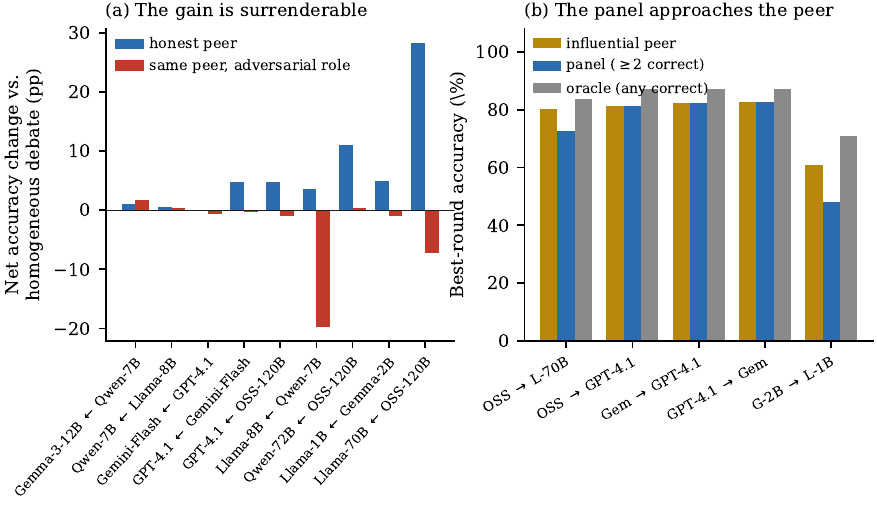}
\caption{Borrowed benefit and imported risk across the nine directions. Left:
direction-specific controlled accuracy under an honest peer and the same peer
under an adversarial role. Right: in the five full-reference settings, the best
majority correctness does not exceed the influential peer's best observed accuracy while the any-member
oracle is higher.}
\label{fig:surrender}
\end{figure*}

\section{Confirmatory Fixed-Qwen Details}
\label{app:qwen-confirmatory}

All three fixed-receiver conditions have 786 matched replicate-items at R1. The
controlled GPT-OSS-minus-Gemma gain difference is +9.03 points with paired
benchmark-item bootstrap interval [5.14,12.88]. The reception difference is
$-1.81$ points [$-13.01$,9.87], while adversarial uptake increases by +3.41
points [1.12,5.82]. At R4, honest Qwen receivers reach 69.28\% with Gemma and
79.20\% with GPT-OSS; the respective homogeneous Qwen support is approximately
59\%. In the GPT-OSS panel, the peer itself peaks at 80.62\%, majority
correctness at 79.97\%, and the any-member oracle at 85.37\%.

The item-level recovery trend pools Qwen receiver exposures that begin wrong in
$[Q,Q,Q]$, $[Q,Q,G]$, and $[Q,Q,O]$, with the shared homogeneous baseline entered once rather than once per rung. A logistic model includes peer gap per ten
points, replicate indicators, and receiver-slot indicator, with a benchmark-item
clustered sandwich variance. It yields OR 3.03 [2.47,3.73]. The analogous model
on R0-correct exposures yields honest-corruption OR .98 [.72,1.34]. These errors
condition on this benchmark's items and do not turn three peer levels into a
population of models.

\section{Trajectory and Ceiling Details}
\label{app:trajectories}

Table~\ref{tab:headroom} reports per-direction headroom and the fraction of it
closed; Table~\ref{tab:rounds} gives the round-by-round accuracy curves.

\begin{table*}[t]
\centering
\small
\begin{tabular}{lrrrrrr}
\toprule
Direction & Rep.-items & Recv.\ $A_0$ & Peer $A_0$ & Mixed best & Controlled peak (pp) & Gap closed \\
\midrule
Gemma-3-12B $\leftarrow$ Qwen-7B & 715 & 72.0 & 63.2 & 73.2 (R2) & +2.0 & -- \\
Qwen-7B $\leftarrow$ Llama-8B & 3957 & 91.9 & 87.1 & 92.9 (R3) & +0.7 & -- \\
Gemini-Flash $\leftarrow$ GPT-4.1 & 784 & 79.5 & 75.9 & 82.2 (R1) & +0.7 & -- \\
GPT-4.1 $\leftarrow$ Gemini-Flash & 782 & 75.8 & 79.6 & 82.2 (R3) & +7.6 & 167\% \\
GPT-4.1 $\leftarrow$ OSS-120B & 784 & 75.9 & 80.1 & 81.3 (R4) & +6.4 & 127.5\% \\
Llama-8B $\leftarrow$ Qwen-7B & 3957 & 87.1 & 91.9 & 91.5 (R4) & +4.5 & 91\% \\
Qwen-72B $\leftarrow$ OSS-120B & 763 & 67.3 & 81.5 & 81.3 (R4) & +14.0 & 99\% \\
Llama-1B $\leftarrow$ Gemma-2B & 100 & 37.7 & 61.0 & 43.5 (R1) & +8.8 & 25\% \\
Llama-70B $\leftarrow$ OSS-120B & 745 & 46.7 & 80.5 & 72.7 (R3) & +30.2 & 77\% \\
\bottomrule
\end{tabular}
\caption{Full-trajectory view of the honest panels. This analysis uses its own, stricter matched support: a replicate-item is retained only when every round R0--R4 is complete in both the homogeneous and mixed panels, so cluster counts, $A_0$, and $\Gamma$ differ slightly from Table~\ref{tab:directions}, where matching is on the R0/R1 cells the transition estimand needs. \emph{Controlled peak} subtracts the receiver's homogeneous accuracy at the same round, removing self-panel drift. \emph{Gap closed} is the controlled peak as a fraction of $\Gamma$, defined only where $\Gamma > 0$; values above 100\% mean the receivers ended above the measured gap, which is possible because the peer also gains from the exchange.}
\label{tab:headroom}
\end{table*}

\begin{table*}[t]
\centering
\small
\begin{tabular}{llrrrrr}
\toprule
Setting & Series & R0 & R1 & R2 & R3 & R4 \\
\midrule
GPT-OSS-120B $\to$ Llama-70B & Peer alone & 80.5 & 81.1 & 81.1 & 80.9 & 80.8 \\
 & Peer in panel & 80.4 & 79.4 & 79.4 & 78.1 & 77.3 \\
 & Receivers & 44.3 & 70.7 & 72.1 & 72.0 & 71.1 \\
 & Panel $\geq$2 & 45.2 & 71.8 & 72.8 & 72.7 & 71.6 \\
 & Oracle & 82.5 & 83.6 & 82.9 & 81.7 & 80.3 \\
\midrule
GPT-OSS-120B $\to$ GPT-4.1 & Peer alone & 80.2 & 80.8 & 80.8 & 80.7 & 80.6 \\
 & Peer in panel & 79.4 & 81.4 & 79.1 & 78.1 & 77.7 \\
 & Receivers & 75.9 & 80.4 & 80.9 & 81.0 & 81.4 \\
 & Panel $\geq$2 & 77.6 & 81.4 & 81.2 & 81.0 & 81.3 \\
 & Oracle & 87.2 & 85.6 & 83.2 & 82.2 & 81.8 \\
\midrule
Gemini-2.5-Flash $\to$ GPT-4.1 & Peer alone & 79.5 & 81.6 & 81.3 & 81.4 & 81.3 \\
 & Peer in panel & 79.7 & 82.5 & 81.0 & 80.4 & 80.4 \\
 & Receivers & 77.1 & 81.6 & 82.0 & 82.2 & 82.2 \\
 & Panel $\geq$2 & 79.2 & 82.3 & 82.5 & 82.1 & 82.3 \\
 & Oracle & 87.2 & 87.1 & 84.6 & 84.1 & 83.9 \\
\midrule
GPT-4.1 $\to$ Gemini-2.5-Flash & Peer alone & 75.9 & 75.5 & 74.4 & 74.7 & 74.9 \\
 & Peer in panel & 77.0 & 82.5 & 82.7 & 82.5 & 82.5 \\
 & Receivers & 80.1 & 82.2 & 81.4 & 80.9 & 80.9 \\
 & Panel $\geq$2 & 80.5 & 82.7 & 81.6 & 81.1 & 81.1 \\
 & Oracle & 87.4 & 86.9 & 84.8 & 84.1 & 84.1 \\
\midrule
Gemma-2B $\to$ Llama-1B & Peer alone & 61.0 & -- & -- & -- & -- \\
 & Peer in panel & 61.0 & 54.0 & 55.0 & 53.0 & 49.0 \\
 & Receivers & 41.5 & 43.5 & 43.5 & 42.5 & 41.5 \\
 & Panel $\geq$2 & 42.0 & 46.0 & 48.0 & 46.0 & 42.0 \\
 & Oracle & 71.0 & 61.0 & 58.0 & 56.0 & 56.0 \\
\bottomrule
\end{tabular}
\caption{Round-by-round accuracy in the five full-reference settings. In all five, the peer's own accuracy inside the panel declines from its best round to the final round --- most sharply for Gemma-2B beside two Llama-1B receivers, which falls from 61.0\% to 49.0\%. A best-round comparison alone would hide this. Missing cells are rounds absent from the small-model diagnostic's standalone trace, which has one agent and one round.}
\label{tab:rounds}
\end{table*}

\begin{table*}[t]
\centering
\small
\begin{tabular}{lrrrrrrr}
\toprule
& & \multicolumn{2}{c}{Peer (\%)} & \multicolumn{3}{c}{Panel (\%)} & \\
\cmidrule(lr){3-4}\cmidrule(lr){5-7}
Setting & $n$ & alone & in panel & receivers & $\geq$2 & oracle & Shortfall \\
\midrule
GPT-OSS-120B $\to$ Llama-70B & 776 & 81.1 & 80.4 & 72.1 & 72.8 & 83.6 & $-$10.8 \\
GPT-OSS-120B $\to$ GPT-4.1 & 786 & 80.8 & 81.4 & 81.4 & 81.4 & 87.2 & $-$5.7 \\
Gemini-2.5-Flash $\to$ GPT-4.1 & 784 & 81.6 & 82.5 & 82.2 & 82.5 & 87.2 & $-$4.7 \\
GPT-4.1 $\to$ Gemini-2.5-Flash & 784 & 75.9 & 82.7 & 82.2 & 82.7 & 87.4 & $-$4.7 \\
Gemma-2B $\to$ Llama-1B & 100 & 61.0 & 61.0 & 43.5 & 48.0 & 71.0 & $-$23.0 \\
\bottomrule
\end{tabular}
\caption{Where the panel stops. Accuracies are best-round values on $n$ matched
replicate-items. \emph{Peer alone} is that model's best round in its own
homogeneous panel; \emph{peer in panel} its best in slot 2 of the honest
heterogeneous panel; \emph{$\geq$2} requires two of three agents correct;
\emph{oracle} any one; \emph{shortfall} is $\geq$2 minus oracle. Maxima are
selected after observing the rounds; round-0 comparisons and full curves are in
the appendix.}
\label{tab:ceiling}
\end{table*}

\begin{table*}[t]
\centering
\scriptsize
\setlength{\tabcolsep}{4pt}
\begin{tabular}{lrrrrrrrr}
\toprule
Setting & $n$ & Panel & Peer & Panel-only & Peer-only & Net & New & Discarded \\
\midrule
GPT-OSS-120B $\to$ Llama-70B & 776 & 71.65 & 77.32 & 2.96 & 8.63 & $-$5.67 & 2.58 & 13.40 \\
GPT-OSS-120B $\to$ GPT-4.1 & 786 & 81.30 & 77.74 & 3.69 & 0.13 & +3.56 & 1.65 & 7.51 \\
Gemini-2.5-Flash $\to$ GPT-4.1 & 784 & 82.27 & 80.36 & 3.19 & 1.28 & +1.91 & 1.91 & 6.89 \\
GPT-4.1 $\to$ Gemini-2.5-Flash & 784 & 81.12 & 82.53 & 1.53 & 2.93 & $-$1.40 & 2.04 & 8.29 \\
Gemma-2B $\to$ Llama-1B & 100 & 42.00 & 49.00 & 1.00 & 8.00 & $-$7.00 & 1.00 & 30.00 \\
\bottomrule
\end{tabular}
\caption{Fixed-round item accounting in the five full-reference honest panels
at R4 (\%). \emph{Panel} requires at least two correct members; \emph{Peer} is
the influential peer in that panel. \emph{Panel-only} and \emph{peer-only} are
the two discordant outcomes, with \emph{Net} equal to their difference.
\emph{New} denotes a correct R4 panel when no member was correct at R0;
\emph{discarded} denotes an incorrect R4 panel despite at least one correct R0
member. New outcomes demonstrate item-level exceptions to a pure relay account,
but do not by themselves identify a causal composition mechanism. The fixed
round avoids selecting these contrasts after observing their values.}
\label{tab:composition-accounting}
\end{table*}

The Qwen--GPT-OSS setting follows the pattern of the five full-reference
settings: the at-least-two-correct panel peaks at 79.97\%, the peer in panel at
80.62\%, and the oracle at 85.37\%.

\section{Scope Controls}
\label{app:controls}

\subsection{Open-answer multi-hop QA}

The MuSiQue cohort contains six complete panels on the same 300 questions,
balanced across two-, three-, and four-hop items: $[G,G,G]$, $[Q,Q,Q]$,
$[G,G,Q]$, $[G,G,Q^{\mathrm{adv}}]$, $[G,G,G^{\mathrm{adv}}]$, and
$[G,G^{\mathrm{adv}},Q]$. Answers are scored against references and aliases
using normalized exact match and token F1; no LLM judge is used. The
Gemma-homogeneous round curve remains useful as a task baseline: pooled exact
match rises from 45.67\% at R0 to 48.11\% at R1 and ends at 47.44\%
(Table~\ref{tab:musique}).

\begin{table}[t]
\centering
\small
\begin{tabular}{lrrrr}
\toprule
Round & Pooled EM & $\geq$2 & Oracle & F1 \\
\midrule
R0 & 45.67 & 45.67 & 47.33 & 58.90 \\
R1 & 48.11 & 48.33 & 49.00 & 61.86 \\
R2 & 48.00 & 48.00 & 49.33 & 61.27 \\
R3 & 48.00 & 48.33 & 48.67 & 61.00 \\
R4 & 47.44 & 47.33 & 48.33 & 60.57 \\
\bottomrule
\end{tabular}
\caption{Homogeneous baseline within the MuSiQue cohort: 300 items, one Gemma-3-12B panel, and alias-normalized exact match recomputed from raw answers with no LLM judge. Pooled exact match rises 2.44~pp to a peak at R1 and gives back 0.67~pp by R4. The oracle exceeds realized consensus by at most 1.67~pp here, against 4.7--23.0~pp in the five mathematics settings of Table~\ref{tab:ceiling}. The matched heterogeneous comparison is in Table~\ref{tab:musique-heterogeneous}.}
\label{tab:musique}
\end{table}

The heterogeneous comparison supplies the cross-task scope probe. Qwen's R0
capability is 10.78 points below Gemma's. At R1, an honest Qwen peer raises
recovery and corruption by nearly equal amounts, leaving controlled net gain at
+1.17 points; reception is nevertheless 62.50\%. The adversarial Qwen peer adds
2.66 points of excess corruption at R1 and reaches a controlled $-4.50$ points
by R4. This is an ungated direction: active reception with negative aggregate
peer advantage produces
little first-exchange gain or leverage, while repeated adversarial influence can
still accumulate.

Two further control panels separate that heterogeneous result from broader threat claims. A same-family Gemma adversary raises R1 corruption by 8.47 points and
reaches $-9.50$ controlled points at R4. In the contaminated panel, an honest Qwen peer improves the sole honest Gemma receiver by +5.00 controlled
points at R1 [1.02,9.18], but the R4 effect is $-3.00$ [$-8.36$,2.49]. This is a single run, and we do not generalize from it.

\begin{table*}[t]
\centering
\small
\setlength{\tabcolsep}{4pt}
\begin{tabular}{lrrr}
\toprule
Quantity & Estimate & Bootstrap 95\% CI & $n$ \\
\midrule
Capability gap $\Gamma$ (pp) & $-$10.78 & [$-$16.54, $-$4.79] & 300 \\
Honest controlled gain, R1 (pp) & +1.17 & [$-$2.61, +4.95] & 300 \\
Honest recovery difference, R1 (pp) & +9.67 & [+4.78, +14.84] & 326 \\
Honest corruption difference, R1 (pp) & +9.25 & [+4.19, +14.55] & 271 \\
Adversarial-peer controlled gain, R1 (pp) & +1.00 & [$-$1.89, +3.83] & 300 \\
Honest controlled gain, R4 (pp) & +2.33 & [$-$2.08, +6.73] & 300 \\
Adversarial-peer controlled gain, R4 (pp) & $-$4.50 & [$-$8.78, $-$0.32] & 300 \\
Adversarial-peer excess corruption, R1 (pp) & +2.66 & [$-$0.77, +6.45] & 268 \\
Honest reception $\rho$ (\%) & 62.50 & [46.57, 78.13] & 64 \\
Unaided revision $\rho_0$ (\%) & 4.91 & [1.92, 8.43] & 265 \\
Adversarial uptake $\upsilon$ (\%) & 4.85 & [1.77, 8.63] & 268 \\
Same-family adversary excess corruption, R1 (pp) & +8.47 & [+3.45, +13.97] & 272 \\
Same-family adversary controlled gain, R4 (pp) & $-$9.50 & [$-$14.39, $-$4.85] & 300 \\
Honest Qwen effect under contamination, R1 (pp) & +5.00 & [+1.02, +9.18] & 300 \\
Honest Qwen effect under contamination, R4 (pp) & $-$3.00 & [$-$8.36, +2.49] & 300 \\
\bottomrule
\end{tabular}
\caption{Heterogeneous MuSiQue-Ans scope probe on 300 matched open-answer multi-hop questions. Gemma-3-12B is the receiver and Qwen-7B the heterogeneous peer. Controlled gains are difference-in-change against $[G,G,G]$; transition quantities are measured at R0$\rightarrow$R1. The same-family adversary is $[G,G,G^{\mathrm{adv}}]$. The contaminated comparison replaces one honest Gemma in that panel with Qwen, yielding $[G,G^{\mathrm{adv}},Q]$. Intervals are benchmark-item Poisson bootstraps.}
\label{tab:musique-heterogeneous}
\end{table*}

\subsection{Prompt, temperature, and seed variation}

To test whether complementary information can be induced without model
diversity, we compare uniform Gemma-3-12B debate with a same-model panel whose
agents use careful, critical, and concise prompt emphases at temperatures
0, .3, and .7. Both conditions use seeds 0, 1, and 2 on the same 262 MATH-hard
items. The diverse configuration increases R0 any-agent oracle coverage from
67.81\% to 69.97\%, demonstrating more complementary answers. Debate does not
realize them: majority correctness peaks at 63.74\%, below the
uniform panel's 67.94\%, and the diverse-panel oracle contracts by 4.45 points
as agents converge (Table~\ref{tab:agent-diversity-ablation}). This control does not exhaust prompt or sampling diversity;
it shows that eliciting complementarity is insufficient when the protocol does
not preserve and select it.

\begin{table*}[t]
\centering
\small
\begin{tabular}{llrrrrr}
\toprule
Panel & Metric & R0 & R1 & R2 & R3 & R4 \\
\midrule
Diverse config. & Pooled agent & 60.31 & 63.32 & 63.40 & 63.36 & 63.49 \\
 & At least two correct & 60.43 & 63.61 & 63.49 & 63.10 & 63.74 \\
 & Any-agent oracle & 69.97 & 68.07 & 66.79 & 66.16 & 65.52 \\
\addlinespace
Uniform config. & Pooled agent & 67.81 & 68.07 & 67.13 & 66.45 & 66.71 \\
 & At least two correct & 67.81 & 67.94 & 67.18 & 66.54 & 66.79 \\
 & Any-agent oracle & 67.81 & 68.45 & 67.30 & 66.79 & 66.92 \\
\bottomrule
\end{tabular}
\caption{Same-model diversity control on MATH-hard (three seeds, 262 items
each). The diverse Gemma-3-12B panel assigns careful/critical/concise prompt
emphases and temperatures 0/.3/.7 to its three agents; the uniform panel uses a
shared prompt at temperature 0. Seeds are 0/1/2 in both conditions. Diverse
decoding increases R0 oracle coverage but not realized panel agreement. Parse
failures count as incorrect.}
\label{tab:agent-diversity-ablation}
\end{table*}

\section{Sensitivity and Structural Coupling}
\label{app:sensitivity}

Table~\ref{tab:sensitivity} reports the diagnostic-exclusion sensitivity
analysis, including cross-direction associations and alternative specifications.

\begin{table}[t]
\centering
\small
\begin{tabular}{lrr}
\toprule
Association & $r$ & Exact $p$ \\
\midrule
$r(\Gamma, \text{honest net gain})$ & +0.973 & 0.0009 \\
$r(\Gamma, \text{adversarial corruption rate})$ & +0.729 & 0.0387 \\
$r(\Gamma, \text{corruption vs.\ homogeneous})$ & +0.607 & 0.0914 \\
$r(\Gamma, \rho)$ & +0.549 & 0.1544 \\
$r(H, \rho)$ & +0.113 & 0.7973 \\
$r(\rho, \upsilon)$ & +0.611 & 0.1120 \\
\midrule
Specification & Slope & $R^2$ \\
\midrule
$\Gamma^{+}$ & 0.759 & 0.9925 \\
$\rho$ & 14.197 & 0.1590 \\
$\Gamma^{+} \times \rho$ & 1.087 & 0.9951 \\
\bottomrule
\end{tabular}
\caption{Every cross-direction relation refitted on the eight directions that remain when the 100-item small-model diagnostic is removed, with exact $p$ over all $8! = 40{,}320$ label permutations. No sign changes and no qualitative change: the gate relations strengthen, the two reception null results stay null, and the product specification retains the highest $R^2$ of the three. Computed arithmetically from the frozen per-direction values reported in Tables~\ref{tab:reception} and~\ref{tab:rates}.}
\label{tab:sensitivity}
\end{table}

\begin{table*}[t]
\centering
\scriptsize
\setlength{\tabcolsep}{3.5pt}
\begin{tabular}{lrrrrrrrr}
\toprule
Direction & $\Gamma$ & $S$ & $\rho$ & $S\rho$ & $T$ & $\eta$ & $T\eta$ & Net \\
& \multicolumn{8}{c}{\scriptsize percentage points, except conditional rates $\rho$ and $\eta$ (\%)} \\
\midrule
Gemma-3-12B $\leftarrow$ Qwen-7B & $-$9.7 & 7.4 & 14.7 & 1.08 & 16.4 & 5.8 & 0.95 & +0.13 \\
Qwen-7B $\leftarrow$ Llama-8B & $-$4.8 & 3.1 & 36.9 & 1.14 & 8.6 & 5.7 & 0.49 & +0.64 \\
Gemini-Flash $\leftarrow$ GPT-4.1 & $-$3.5 & 7.2 & 32.5 & 2.35 & 10.2 & 7.5 & 0.76 & +1.59 \\
GPT-4.1 $\leftarrow$ Gemini-Flash & +3.5 & 9.2 & 83.3 & 7.63 & 6.5 & 24.3 & 1.59 & +6.04 \\
GPT-4.1 $\leftarrow$ OSS-120B & +4.3 & 9.8 & 83.1 & 8.14 & 6.3 & 34.3 & 2.16 & +5.98 \\
Llama-8B $\leftarrow$ Qwen-7B & +4.8 & 8.0 & 78.5 & 6.27 & 3.1 & 84.2 & 2.62 & +3.65 \\
Qwen-72B $\leftarrow$ OSS-120B & +14.7 & 16.4 & 63.0 & 10.35 & 1.9 & 10.0 & 0.19 & +10.16 \\
Llama-1B $\leftarrow$ Gemma-2B & +23.3 & 28.0 & 17.9 & 5.00 & 8.5 & 41.2 & 3.50 & +1.50 \\
Llama-70B $\leftarrow$ OSS-120B & +36.0 & 38.1 & 71.0 & 27.04 & 2.0 & 0.0 & 0.00 & +27.04 \\
\midrule
Qwen-7B $\leftarrow$ Gemma-3-12B & +9.1 & 16.5 & 62.2 & 10.24 & 7.5 & 31.4 & 2.35 & +7.89 \\
Qwen-7B $\leftarrow$ OSS-120B & +21.5 & 25.2 & 60.4 & 15.20 & 2.0 & 31.2 & 0.64 & +14.57 \\
\midrule
Gemma-3-12B $\leftarrow$ Qwen-7B (MuSiQue) & $-$10.8 & 10.7 & 62.5 & 6.67 & 21.8 & 23.7 & 5.17 & +1.50 \\
\bottomrule
\end{tabular}
\caption{Discordant-state accounting for honest heterogeneous panels at
R0$\rightarrow$R1. $S=P(R_0=0,P_0=1)$ is useful peer-only mass and
$T=P(R_0=1,P_0=0)$ is opposing receiver-only mass, both measured in the honest
mixed panel. $\rho=P(R_1=1\mid R_0=0,P_0=1)$ and
$\eta=P(R_1=0\mid R_0=1,P_0=0)$; the realized flows are $S\rho$ and $T\eta$,
and \emph{Net} is their difference. $\Gamma$ is measured separately from the
matched homogeneous reference panels, so condition-specific R0 variation means
it need not equal $S-T$ exactly. The first nine rows are the discovery set, the
next two the fixed-receiver ladder, and the final row the MuSiQue scope probe.}
\label{tab:discordant-flows}
\end{table*}

\begin{table}[t]
\centering
\small
\setlength{\tabcolsep}{3pt}
\begin{tabular}{lrrrr}
\toprule
Predictor & Slope & Intercept & $R^2$ & Max. residual \\
\midrule
$\Gamma^+\rho$ & 1.088 & +0.61 & 0.9950 & 1.07 \\
$(S-T)^+\rho$ & 1.072 & +0.94 & 0.9898 & 1.51 \\
$S\rho$ & 1.088 & $-$1.79 & 0.9680 & 2.30 \\
$S\rho-T\eta$ & 1.027 & +0.08 & 0.9671 & 3.38 \\
\bottomrule
\end{tabular}
\caption{Alternative descriptions of controlled honest gain on the same nine
directions, all fitted with an intercept. $(S-T)$ is the R0 gap inside the
honest mixed panel, whereas $\Gamma$ uses the matched homogeneous references.
Because $\rho$ and $\eta$ include R1 outcomes, these are accounting diagnostics,
not independent predictive models. Replacing aggregate advantage with gross
useful mass or signed realized discordant flow does not improve the descriptive
fit. Residuals are percentage points.}
\label{tab:flow-specifications}
\end{table}

Table~\ref{tab:discordant-flows} separates useful peer-only mass $S$ from
opposing receiver-only mass $T$. All negative-gap directions still contain
useful peer-only answers; MuSiQue is the clearest cancellation case, with
$S\rho=6.67$ points of realized correction and $T\eta=5.17$ points of realized
corruption within the discordant states. The gate therefore concerns net gain,
not whether useful information exists at all.

The accounting comparison in Table~\ref{tab:flow-specifications} also checks
whether gross useful mass is a better description of controlled gain. It is not:
$S\rho$, $(S-T)^+\rho$, and $S\rho-T\eta$ all fit less closely than the original
$\Gamma^+\rho$ specification on the same nine directions. This does not make
$\rho$ or $\eta$ independent predictors---both contain R1 outcomes. It shows
that the aggregate reference gap remains the tighter empirical summary rather
than merely standing in for all available peer-only correct answers. We do not
use ratios of $\rho_d$ and $\upsilon_d$: they condition on different initial
states and item populations.

\section{Prompt Structure}
\label{app:prompts}

At R0, agents receive the benchmark problem and an instruction to explain their
reasoning and provide a parseable final answer. At R1--R4, each agent receives
the other two complete previous-round responses and is asked to answer the
original problem again. The adversarial peer instead receives the gold answer
and a plausible incorrect target, is prohibited from returning the gold answer,
and is instructed to defend the same target concisely across rounds. The full
verbatim templates are included in the released artifact. No other condition
differs between the honest and adversarial mixed panels.

\end{document}